\documentclass{article}

\usepackage{arxiv}

\usepackage[utf8]{inputenc} % allow utf-8 input
\usepackage[T1]{fontenc}    % use 8-bit T1 fonts
\usepackage{hyperref}       % hyperlinks
\usepackage{url}            % simple URL typesetting
\usepackage{booktabs}       % professional-quality tables
\usepackage{amsfonts}       % blackboard math symbols
\usepackage{nicefrac}       % compact symbols for 1/2, etc.
\usepackage{microtype}      % microtypography
\usepackage{lipsum}
\usepackage{graphicx}

\usepackage{times}
\usepackage{epsfig}
\usepackage{graphicx}
\usepackage{amsmath}
\usepackage{amssymb}
\usepackage{algorithm}
\usepackage{algorithmic}

\title{Hong Kong Air Traffic: Explanation and Prediction based on Sparse Seasonal ARIMA Model}

\author{
  Shuwen Chai \\
  School of Statistics\\
  Renmin University of China\\
  \texttt{chaishuwen@ruc.edu.cn}
}

\begin{document}
\maketitle
\begin{abstract}
	The monthly air traffic of a city is a time series with obvious seasonal pattern, and is closely related to the economic situation and social environment of the city. In Hong Kong, for example, July, August and October tend to be the peak season of traffic flow, while there is also a relatively fixed off-season. In the case of a stable social environment, a carefully identified and fitted seasonal ARIMA model can predict the traffic flow in the future months well. This work selects the air traffic data, including arrival and departure passengers of Hong Kong, after the financial crisis and before the political storm. A sparse seasonal ARIMA$(0,1,1)\times(4,1,0)_{12}$ is built, which can correctly predict the air traffic from January to July in 2020 within its $95\%$ confidence interval. Furthermore, this work decompose the time-series and find that important events, like financial crisis, political storm, and the COVID-19 outbreak, affect the level of air traffic to some extent. For example, the political storm and epidemic prevention and control happened after 2019 made the air traffic drop significantly. According to my sparse seasonal ARIMA model, the air traffic from February to November in 2020 is only $5\%$ of what it should be without these two events. This is a valuable application of time-series model in the air traffic loss estimation.
\end{abstract}

% keywords can be removed

\section{Data Description}

The air traffic data\footnote{Data source: https://www.cad.gov.hk/english/statistics.html} used in this project\footnote{The project is done during exchange semester in HKU.} is obtained from the Civil Aviation Department, consisting of air traffic statistics of the Hong Kong International Airport from 1998 onward. In this project, I ignore the data from 1998 to 2004, since the remaining data is adequate for abundant analysis and the effect of SAS outbreak is similar but less than COVID-19 outbreak.

\begin{figure}[h]
\begin{center}
   \includegraphics[scale=0.5]{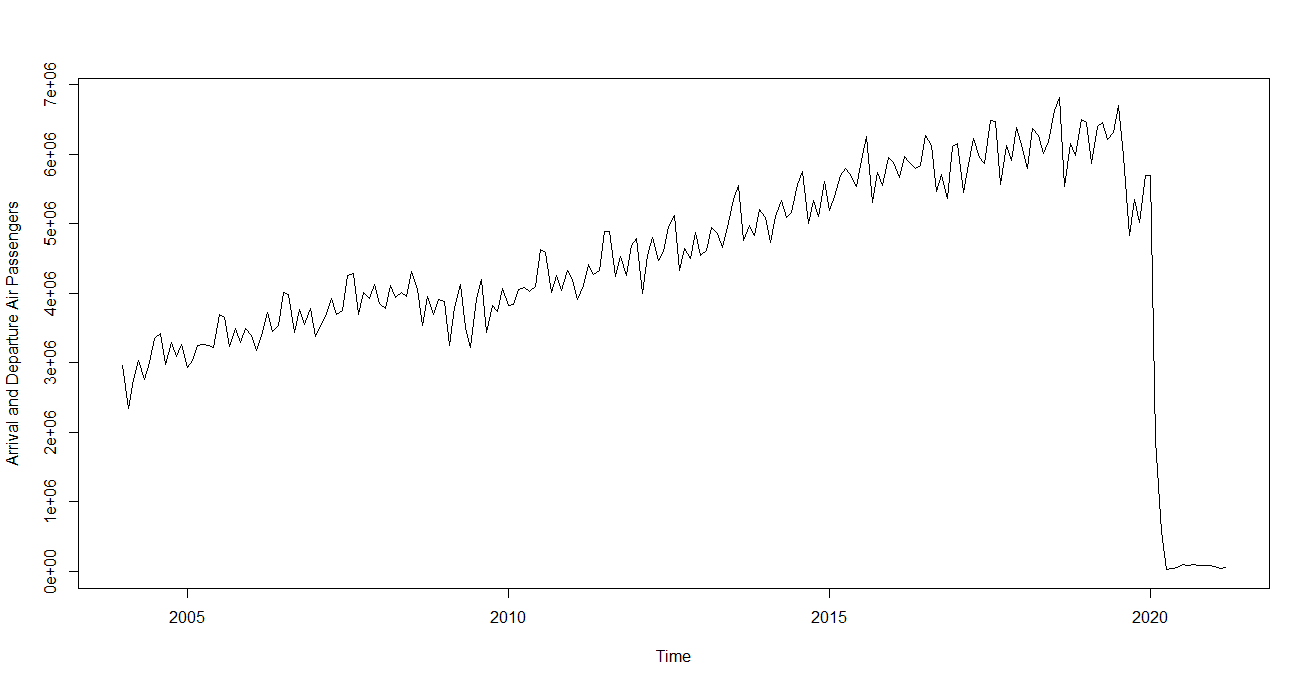}
\end{center}
   \caption{Time Plot of Air Traffic Volume of Hong Kong International Airport}
\label{fig:part1_1_tsall}
\end{figure}

From the time plot \ref{fig:part1_1_tsall} from January 2004 to December 2020, we can observe that the air traffic volume has obvious seasonal pattern\footnote{The seasonality of this time-series is discussed in detail in Section \ref{section:4}.} and increases year by year although fluctuates monthly. Except for the normal observations of seasonality and non-stationarity, there are three subtle anomalies in the time plot.

\begin{itemize}
    \item Firstly, the volume of air traffic dropped a little bit in from early 2008 to 2009, since the financial crisis hit the prosperity of Hong Kong.
    \item The most significant anomaly is that the air traffic dropped to around zero after 2020 due to the COVID-19 outbreak and air traffic control.
    \item Before 2020, there is also a significant drop on the level of air traffic. I guess it is the series of political storms weaken people's willingness to fly to Hong Kong. No matter the actual reason, it serves as another example of social data being significantly affected by a big event (or variable).
\end{itemize}

Considering these unpredictable major social changes, time-series model is imperfect since it focuses simply on the sequence itself, ignoring the other social variables. However, this gives the time-series model another applicable scenario, i.e., to quantifying \ref{section:3} the effect of a major social change on a time-series. The operation is pretty simple. You just need to obtain the forecast of several periods $\{\hat y_{m+1}, \hat y_{m+2}, ..., \hat y_{m+k} \}$ after the impact of major events, and the event influence can be obtained by subtracting the real value $\{y_{m+1}, y_{m+2}, ..., y_{m+k} \}$ from the forecast value. Then, researchers can perform further study based on the difference sequence $\{\Delta_{m+1}, \Delta_{m+2}, ..., \Delta \}$.

The remaining parts of report are arranged as:
\begin{itemize}
    \item In Section \ref{section:2}, I discuss the modelling process for Hong Kong air traffic time-series. After model identification, model diagnosis, and model selection, sparse seasonal $\text{ARIMA}(0,1,1)\times(4,1,0)_{12}$ model is selected. The air traffic in the coming seven months locate precisely in the predicted $95\%$ confidence interval.
    \item In Section \ref{section:3}, the sparse seasonal $\text{ARIMA}(0,1,1)\times(4,1,0)_{12}$ model is applied and fitted again to quantify the negative effect of COVID-19 on air traffic.
    \item Finally, an analysis of seasonality and trend of this time-series is supplemented in Section \ref{section:4}.
\end{itemize}

\section{Seasonal ARIMA Modelling} \label{section:2}
In this section, we use the data from (2009, 1) to (2018, 12) as train data, and the data from (2019, 1) to (2019, 7) as test sequence.

\subsection{Discussion of Stationarity}
\begin{figure}[h]
\begin{center}
   \includegraphics[scale=0.5]{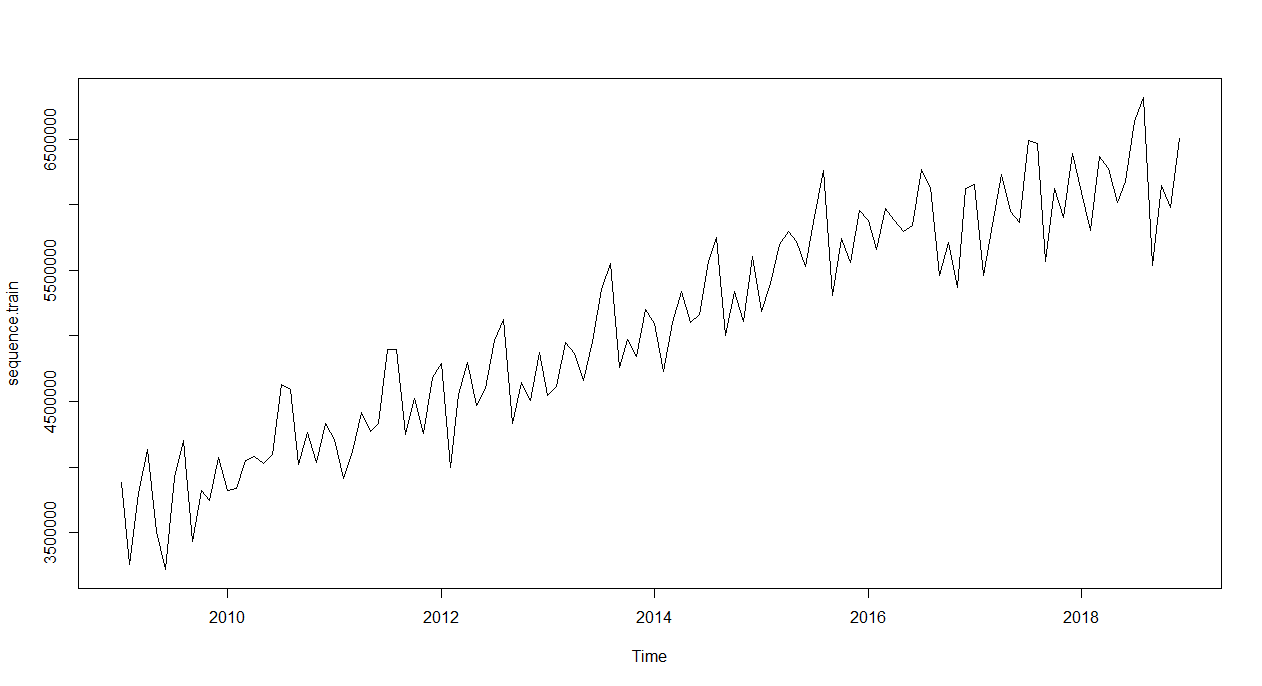}
\end{center}
   \caption{Time plot of air traffic volume of Hong Kong International Airport (Train sequence)}
\label{fig:part1_2_tstrain}
\end{figure}

\begin{figure}[h]
\begin{center}
   \includegraphics[scale=0.4]{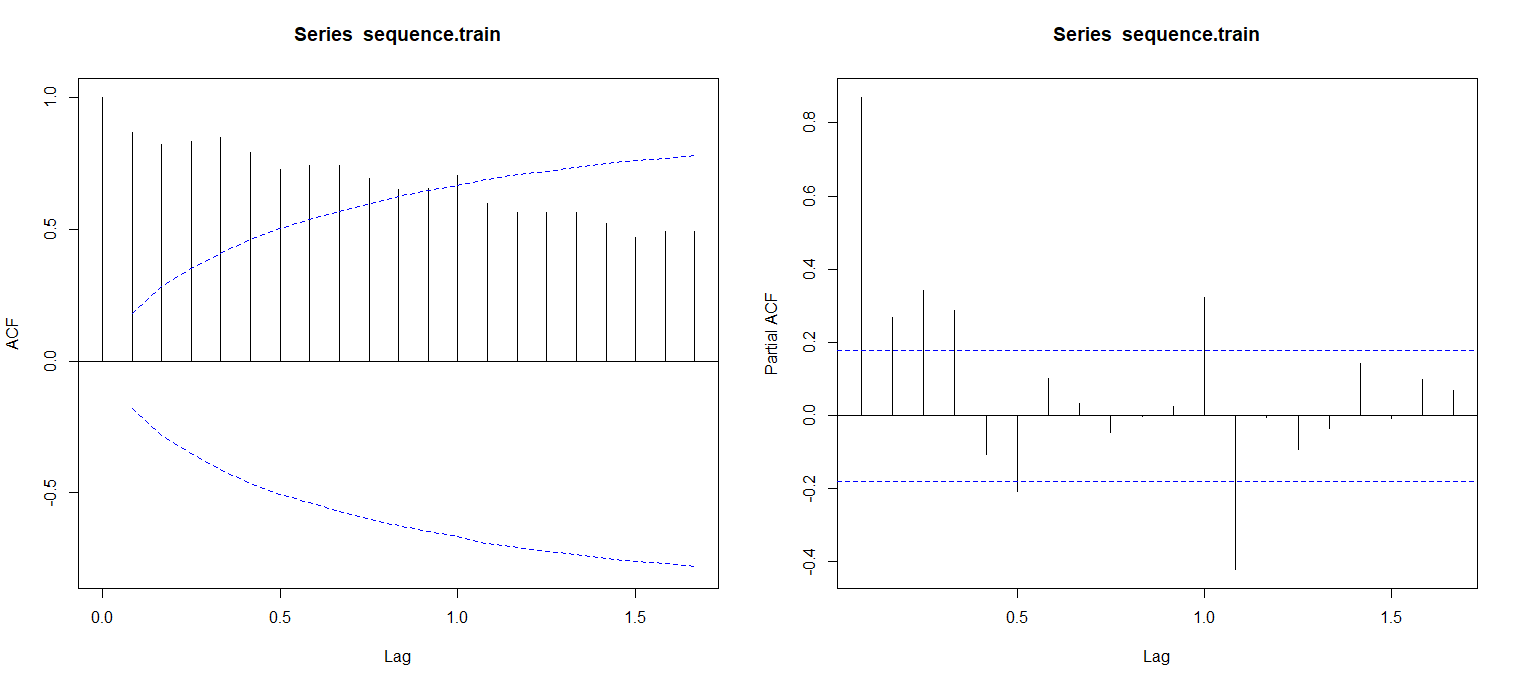}
\end{center}
   \caption{Sample ACF and PACF of the train sequence}
\label{fig:part1_3_acfpacf}
\end{figure}

 To get an initial understanding about the train sequence, we draw the time plot, sample auto-correlation function (ACF), and sample partial auto-correlation function (PACF) as Figure \ref{fig:part1_2_tstrain} and Figure \ref{fig:part1_3_acfpacf}.
 
 We can observe from the time plot that the volume of air passengers went ups and downs in each year, but increases overtime. The sample ACF decays slowly from lag to lag, also indicating that the time-series is non-stationary. Another powerful tool to confirm the non-stationarity is performing the augmented dickey-fuller test (ADF test). 
 
 \begin{equation}
 H_0: \text{The time-series is non stationary} \quad v.s. \quad H_1: \text{The time-series is stationary}
 \end{equation}\label{equ:hypothesis}

 The test result is summarized in table \ref{table:ADF_original}.
 
\begin{table}[h]
	\caption{ADF tests for the original train sequence}
	\centering
	\begin{tabular}{*{9}{c}}
		\toprule
		\multicolumn{3}{c}{Type 1} &
		\multicolumn{3}{c}{Type 2} &
		\multicolumn{3}{c}{Type 3} \\
		\cmidrule(r){1-3} \cmidrule(r){4-6} \cmidrule(r){7-9}
		Lag  & ADF & P.value & Lag  & ADF & P.value & Lag  & ADF & P.value \\
		\midrule
		0 & 0.196 & 0.700 & 0 & -2.309 & 0.208 & 0 & -9.99 & $\leqslant$ 0.01 \\
		1 & 0.753 & 0.859 & 1 & -1.787 & 0.411 & 1 & -7.75 & $\leqslant$ 0.01 \\
		2 & 1.485 & 0.964 & 2 & -1.162 & 0.640 & 2 & -5.01 & $\leqslant$ 0.01 \\
		3 & 2.237 & 0.990 & 3 & -0.951 & 0.714 & 3 & -3.36 & 0.0624 \\
		4 & 2.484 & 0.990 & 4 & -0.955 & 0.712 & 4 & -3.09 & 0.1191 \\
		\bottomrule
	\end{tabular}
	\label{table:ADF_original}
\end{table}

Generally we consider only two types of ADF tests:
 \begin{itemize}
     \item \textbf{Type 1: }, Assuming that the time-series has no drift and no trend.
     \item \textbf{Type 2: }, Assuming that the time-series has drift but no trend. 
 \end{itemize}
 
Actually, the \textbf{Type 3} ADF test assumes that the time-series has drift and trend. Since we are detecting the property of non-stationarity, we should not ignore the trend of the sequence. Thus, by looking at the first two types of ADF tests, the null hypothesis is not rejected consistently, which confirms the non-stationarity. 

Considering the seasonal pattern and increase trend\footnote{See Section \ref{section:4} for detailed analysis.}, we should take the differencing and seasonal differencing both\footnote{I tried to take the differencing only, the differenced sequence could pass the ADF tests but perform poorly in forecasting.}.

Now we can draw the time plot, sample ACF, and sample PACF of the differenced and seasonal differenced train sequence (dsd train sequence). Note that the sequence lost 13 observations after taking these two operations.

\begin{figure}[h]
\begin{center}
   \includegraphics[scale=0.4]{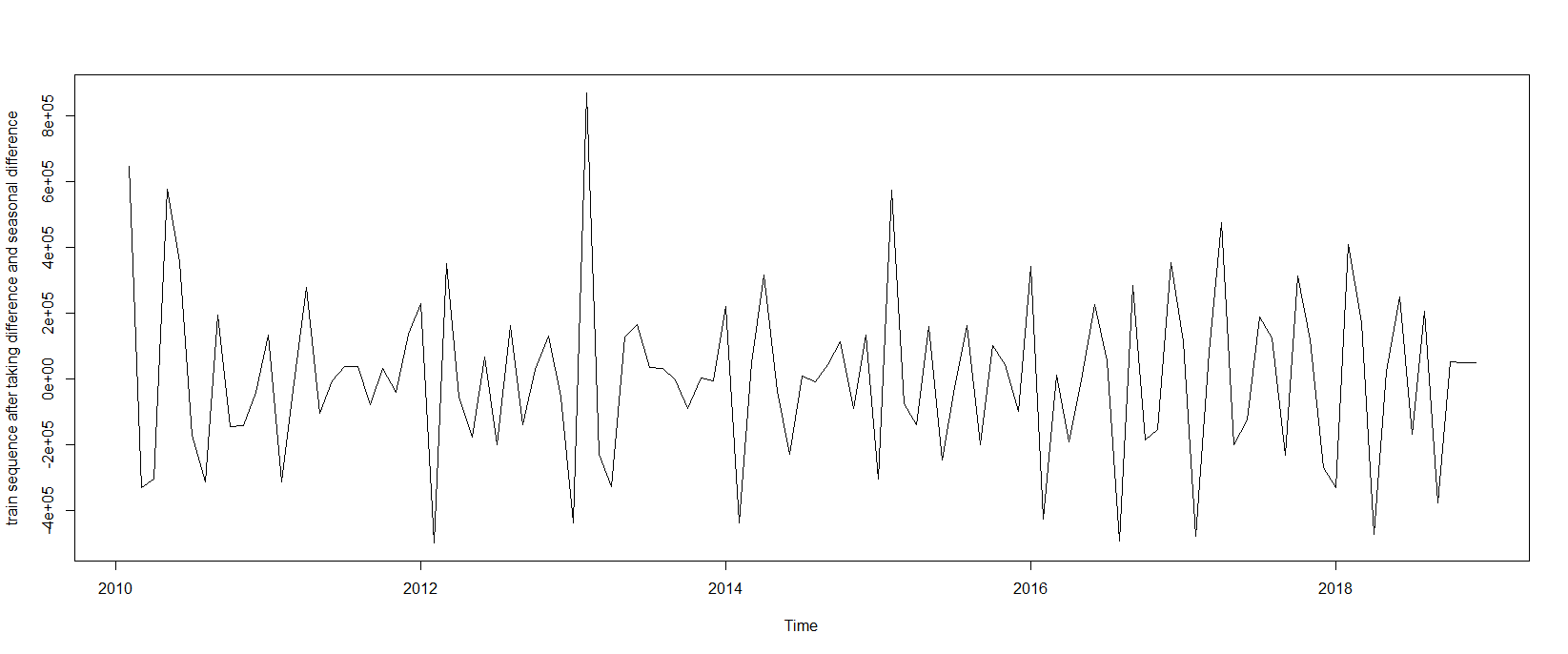}
\end{center}
   \caption{Time plot of dsd train sequence}
\label{fig:part1_4_tstraindsd}
\end{figure}

\begin{figure}[h]
\begin{center}
   \includegraphics[scale=0.38]{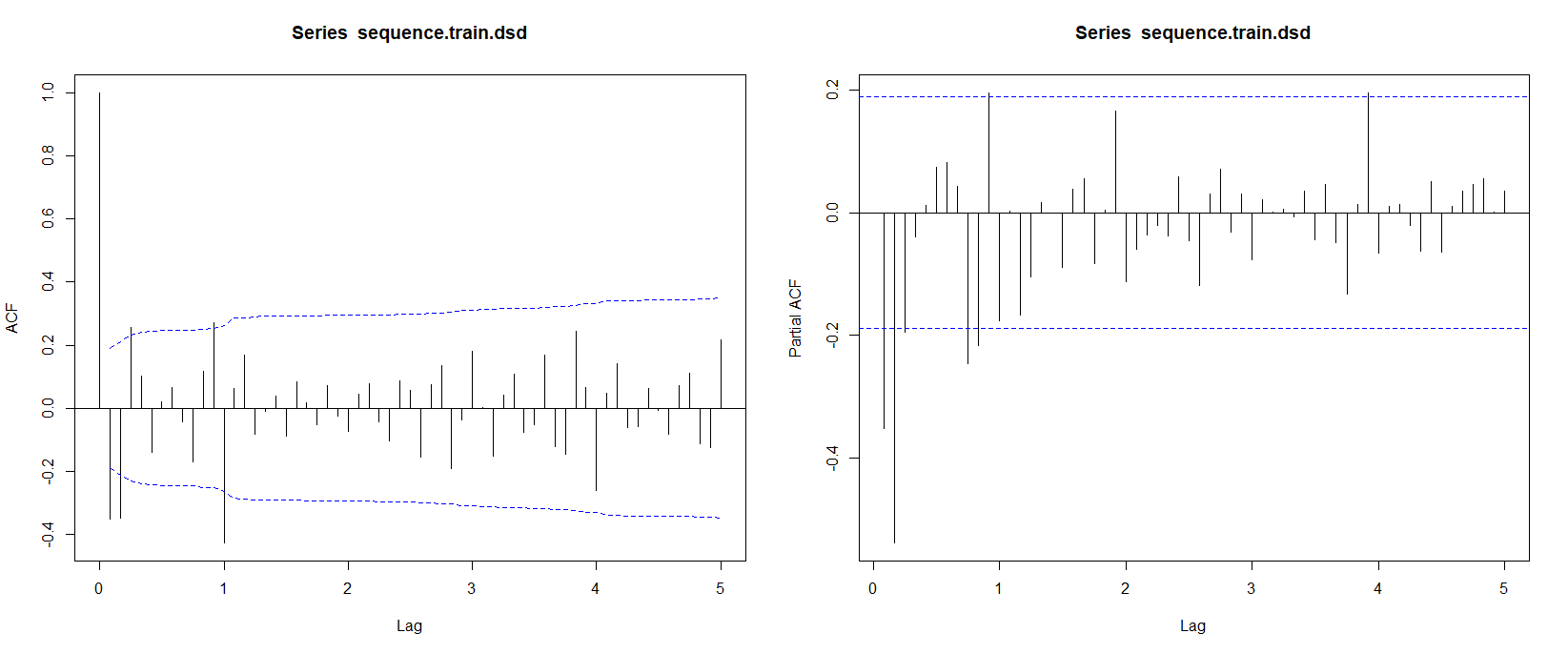}
\end{center}
   \caption{Sample ACF and PACF after taking difference and seasonal difference}
\label{fig:part1_5_acfpacf}
\end{figure}

This time plot \ref{fig:part1_4_tstraindsd} seems contain no trend and fluctuates statioanrily. Then, I perform the ADF test again. The outcome of three types of ADF tests is shown in table \ref{table:ADF_dsd}, with null hypothesis rejected consistently. Note that the sequence after differencing twice will not contain drift, so we do not need to include drift in the model fitting step in the future.

\begin{table}[h]
	\caption{ADF tests for the dsd train sequence}
	\centering
	\begin{tabular}{*{9}{c}}
		\toprule
		\multicolumn{3}{c}{Type 1} &
		\multicolumn{3}{c}{Type 2} &
		\multicolumn{3}{c}{Type 3} \\
		\cmidrule(r){1-3} \cmidrule(r){4-6} \cmidrule(r){7-9}
		Lag  & ADF & P.value & Lag  & ADF & P.value & Lag  & ADF & P.value \\
		\midrule
		0 & -19.26 & $\leqslant$ 0.01 & 0 & -19.20 & $\leqslant$ 0.01 & 0 & -19.14 & $\leqslant$ 0.01 \\
		1 & -19.17 & $\leqslant$ 0.01 & 1 & -19.12 & $\leqslant$ 0.01 & 1 & -19.06 & $\leqslant$ 0.01 \\
		2 & -13.00 & $\leqslant$ 0.01 & 2 & -12.97 & $\leqslant$ 0.01 & 2 & -12.93 & $\leqslant$ 0.01 \\
		3 & -9.17 & $\leqslant$ 0.01 & 3 & -9.15 & $\leqslant$ 0.01 & 3 & -9.12 & $\leqslant$ 0.01 \\
		4 & -7.41 & $\leqslant$ 0.01 & 4 & -7.39 & $\leqslant$ 0.01 & 4 & -7.36 & $\leqslant$ 0.01 \\
		\bottomrule
	\end{tabular}
	\label{table:ADF_dsd}
\end{table}

\subsection{Model Identification}

In this section, I will build a suitable seasonal ARIMA model by specifying the hyper-parameters $p,q,P,Q$ of $\text{ARIMA}(p,1,q)\times(P,1,Q)_{12}$. The sample ACF and PACF plots are shown in Figure \ref{fig:part1_5_acfpacf}.

\subsubsection{Identify Possible Models}

From the first plot, lag-1, lag-2 are significantly outstanding, indicating that the MA part has $q=2$. Lag-12 (seasonal lag-1) is even more outstanding and indicates $Q=1$ for SMA part.

As for the PACF plot, although it is generally considered to be treaky, I believe it can convey lots of useful information. As we can see, a clear spike appears at lag-1. And the partial auto-correlations around seasonal lag 1, 2, 4 locate at the boundary. This pattern of PACF at seasonal lags can not be ignored since it is weird and strong. I believe it is caused by SAR part, which might have hyper-parameter $1 \leqslant P \leqslant 4$. If $P \geqslant 1$ holds, the assumption of $Q=1$ for SMA part might not be true, since SAR will cause the auto-correlations at seasonal lags to decay slowly.

Since the situation is rather complex, I'd like to choose several pairs of possible hyper-parameters for my candidate models, and then fit and test them for adequacy. The four candidate models are summarized in table \ref{table:candidate_model_1}. My choice covers simple model like $\text{Model0}$ with $P=1$ and complex model like Model1 with $P=4$. Since the correlations in the third seasonal lag of PACF plot is not significant, I also consider the Sparse Seasonal ARIMA ($\text{Model3}$), setting the parameter of SAR3 zero constantly. The reason why $\text{Model2}$ and $\text{Model3}$ are verified from $\text{Model1}$ while $\text{Model0}$ has no variations will be revealed later.

\begin{table}[h]
\begin{center}
  \begin{tabular}{c|*{8}c}
  \toprule
  Model Name & Type & p & d & q & P & D & Q & seasonal period \\
  \midrule
  Model0 & Seasonal ARIMA & 0 & 1 & 2 & 1 & 1 & 0 & 12 \\
  Model1 & Seasonal ARIMA & 0 & 1 & 2 & 4 & 1 & 0 & 12 \\
  Model2 & Seasonal ARIMA & 0 & 1 & 1 & 4 & 1 & 0 & 12 \\
  Model3 & Sparse Seasonal ARIMA & 0 & 1 & 1 & 4(3) & 1 & 0 & 12 \\
  \bottomrule
  \end{tabular}
\end{center}
\caption{Four candidate models}
\label{table:candidate_model_1}
\end{table}

\subsubsection{Parameter Estimation for Candidate Models}
The maximum likelihood estimation of the parameters and their corresponding standard errors are provided in table \ref{table:candidates_fitted_1} and table \ref{table:candidates_fitted_2}.

\begin{table}[h]
	\caption{Fitted Model0 and Model1}
	\centering
	\begin{tabular}{*{10}{c}}
		\toprule
		\multicolumn{1}{c}{} &
		\multicolumn{3}{c}{Model0} &
		\multicolumn{6}{c}{Model1} \\
		\cmidrule(r){2-4} 
		\cmidrule(r){5-10}
		& ma1 & ma2 & sar1 & ma1 & ma2 & sar1 & sar2 & sar3 & sar4 \\
		\midrule
        coefficient & -0.6772 & -0.1322 & -0.5195 & -0.6072 & -0.1427 & -0.7524 & -0.5570 & -0.2826 & -0.4239 \\
        standard error & 0.1359 &  \textbf{0.1547} &  0.0914 & 0.1065 &  \textbf{0.1208} & 0.1044 & 0.1447 & \textbf{0.1465} & 0.1099 \\
		\midrule
        & AIC & 2901.09 &  && AIC & 2876.77 & BIC & 2895.48 &\\
		\bottomrule
	\end{tabular}
	\label{table:candidates_fitted_1}
\end{table}

\begin{table}[h]
	\caption{Fitted Model2 and Model3}
	\centering
	\begin{tabular}{*{10}{c}}
		\toprule
		\multicolumn{1}{c}{} &
		\multicolumn{5}{c}{Model2} &
		\multicolumn{4}{c}{Model3} \\
		\cmidrule(r){2-6} 
		\cmidrule(r){7-10}
		& ma1 & sar1 & sar2 & sar3 & sar4 & ma1 & sar1 & sar2 & sar4 \\
		\midrule
        coefficient & -0.6877 & -0.7372 & -0.5510 & -0.2600 & -0.4179 & -0.6960 & -0.6535 & -0.3670 & -0.2897\\
        standard error & 0.0891 & 0.1066 & 0.1473 & 0.1464 & 0.1123 & 0.0827 & 0.0991 & 0.1032 & 0.0940 \\
        \midrule
        & AIC & 2876.13 & BIC & 2892.17 && AIC & 2877.09 & BIC & 2890.46\\
		\bottomrule
	\end{tabular}
	\label{table:candidates_fitted_2}
\end{table}

From the first MLE table \ref{table:candidates_fitted_1}, we can find that the coefficients of ma2 in both Model0 and Model1 are insignificant. Also, the coefficient of sar3 is insignificant. 

Model2 in table \ref{table:candidates_fitted_2} deletes the ma2, but the coefficient of sar3 is still insignificant. Sparse $\text{Model3}$ set the coefficient of sar3 to zero and its remaining coefficients are all significant. Consequently, the Sparse Model3 seems to be a satisfying variation from Model1 up to now.

Note that the reason why I haven't fit a $\text{ARIMA}(0,1,1)\times(1,1,0)_{12}$ model here is that Model0 itself is inadequate. With an inadequate model, it's unwise to make it even simpler, so I will use Model3 for further exploration. For the sack of illustration, I will also show the testing results of Model0 in the following section.

\subsection{Model Diagnosis}

\subsubsection{Residual Analysis of Candidate Models}
Residual analysis will be performed to check the residual normality and model adequacy. 

\begin{figure}[h]
\begin{center}
   \includegraphics[scale=0.35]{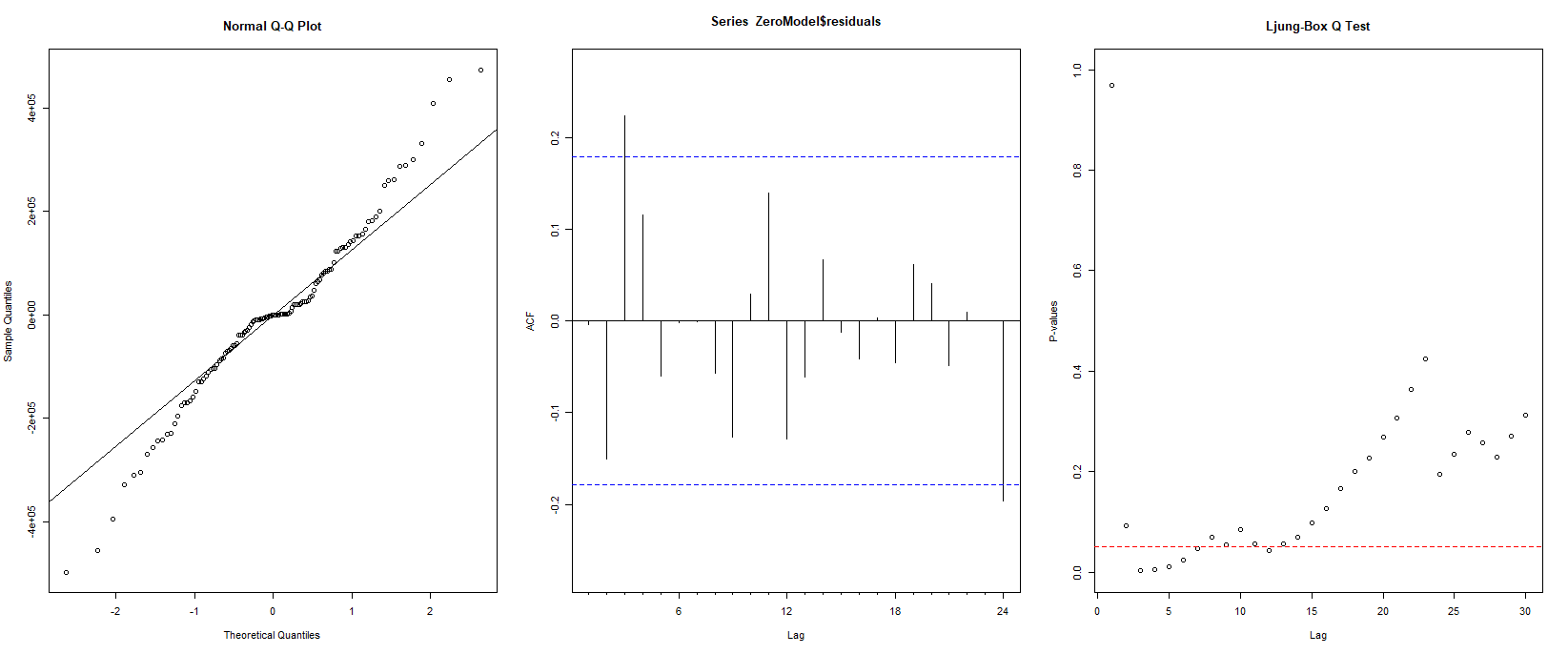}
\end{center}
   \caption{Residual Analysis of Model0}
\label{fig:part1_6_residualModel0}
\end{figure}

\begin{figure}[h]
\begin{center}
   \includegraphics[scale=0.35]{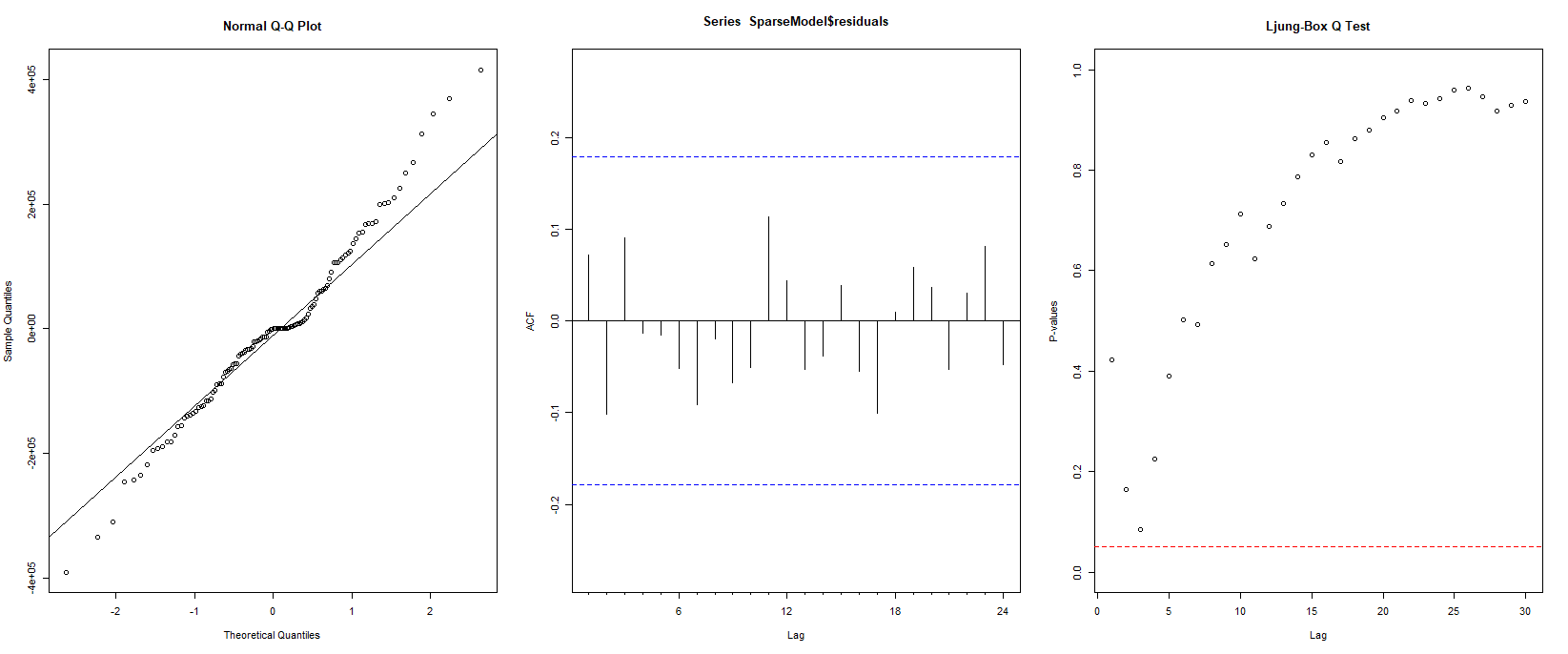}
\end{center}
   \caption{Residual Analysis of Model3}
\label{fig:part1_7_residualModel3}
\end{figure}

The Ljung-box test in Figure \ref{fig:part1_6_residualModel0} for Model0 shows that many test statistics at different lags have value less than the threshold 0.05. Whats more, the Shapiro-Wilk normality test reject the normal hypothesis with a p-value of 0.01871.

As for the Model3, in Figure \ref{fig:part1_7_residualModel3}, the residuals locate near the 45 degree line and the p-value of Shapiro-Wilk Normality Test is 0.1006, p-value of Jarque Bera Test is 0.1683, both greater than 0.05. Combining tests and observation, we don’t reject the normal hypothesis. On the other hand, the Ljung-box test and the ACF show that there is no other correlation pattern left in the residuals, thus Model3 is adequate and promising enough.

\subsubsection{Analysis of Over-Parameterized Models}
To check the stability of Model3, we will fit some over-parameterized models for comparison. Since increasing the order of MA reaches to Model2, which has insignificant coefficient of ma2. Increasing the order of AR brings calculation problem for likelihood, making MLE, AIC, AICc, and BIC incalculable. So here we just consider two cases of over-parametrized models \footnote{Both are sparse models with sar3=0.}:
$$
\text{OverSAR}: \text{ARIMA}(0,1,1)\times(5,1,0)_{12} \quad \text{OverSMA}: \text{ARIMA}(0,1,1)\times(4,1,1)_{12}
$$

The MLE of $\text{OverSAR}$ and $\text{OverSAR}$ are summarized in table \ref{table:overfitted_1}.

\begin{table}[h]
	\caption{MLE of over-parameterized models}
	\centering
	\begin{tabular}{*{11}{c}}
		\toprule
		\multicolumn{1}{c}{} &
		\multicolumn{5}{c}{OverSAR} &
		\multicolumn{5}{c}{OverSMA} \\
		\cmidrule(r){2-6} 
		\cmidrule(r){7-11}
		& ma1 & sar1 & sar2 & sar4 & sar5 & ma1 & sar1 & sar2 & sar4 & sma1 \\
		\midrule
        coef. & -0.6962 & -0.6619 & -0.3618 & -0.3062 & -0.0323 & -0.6997 & -0.3387 & -0.2191 & -0.3847 & -0.4654 \\
        s.e. & 0.0821 & 0.1031 & 0.1037 &  0.1117 & 0.1197 & 0.0824 & 0.1590  & 0.1362 &  0.1051 & 0.2119 \\
        \midrule
        & AIC & 2879.02 & BIC & 2895.06&& AIC & 2876.2 & BIC & 2892.24\\
		\bottomrule
	\end{tabular}
	\label{table:overfitted_1}
\end{table}

The sar5 for $\text{OverSAR}$ is insignificant, so $\text{Model3}$ is stable. The sma1 for $\text{OverSMA}$ can be regarded as significant while its sar2 becomes insignificant. So we may fit and test the adequacy of another model. $\text{Model4}: \text{ARIMA}(0,1,1)\times(4,1,1)_{12}$ with sar2=0 and sar3=0. The MLE result in table \ref{table:candidates_fitted_3} shows that the coefficient of sar1 is insignificant, then I also fit the $\text{Model5}: \text{ARIMA}(0,1,1)\times(4,1,1)_{12}$ with sar1 = sar2 = sar3 = 0, whose coefficients are all significant.

\begin{table}[h]
	\caption{MLE of Model4 and Model5}
	\centering
	\begin{tabular}{*{10}{c}}
		\toprule
		\multicolumn{1}{c}{} &
		\multicolumn{4}{c}{OverSAR} &
		\multicolumn{4}{c}{OverSMA} \\
		\cmidrule(r){2-5} 
		\cmidrule(r){6-9}
		& ma1 & sar1 & sar4 & sma1 & ma1 & sar4 & sma1 & \\
		\midrule
        coef. & -0.7148 & -0.2018 & -0.3949 & -0.6534 & -0.7251 & -0.3631  & -0.8222 &\\
        s.e. & 0.0782 & 0.1190 &  0.1133 &  0.1571 & 0.0761 & 0.1236 & 0.1667 &\\
        \midrule
        & AIC & 2876.52 & BIC & 2889.89 & AIC & 2877.18 & BIC & 2887.87\\
		\bottomrule
	\end{tabular}
	\label{table:candidates_fitted_3}
\end{table}

Then I carried out the Ljung-box test for $\text{Model5}$, the result shows that this model is not so adequate as $\text{Model3}$. Considering the AIC and BIC of these two models are rather similar, I choose $\text{Model3}$ as my final model.

\subsection{Forecasting}
\subsubsection{Forecast 7 periods}
With my final model, I forecast the air traffic of Hong Kong International Airport from January 2019 to July 2019. All the true value fall in the 95$\%$ confidence interval\footnote{The red line indicates the fitted value of train sequence. The black dotted line indicates the true value. The blue line is the predicted sequence, the shallow shadowed area is the 80$\%$ C.I. and the heavy shadowed area is the 95$\%$ C.I..}. The selected model functions well in predicting air traffic when the social environment is rather stable.

For the convenience of reading, I also put down the number in table \ref{table:forecast7}.

\begin{figure}[h]
\begin{center}
   \includegraphics[scale=0.38]{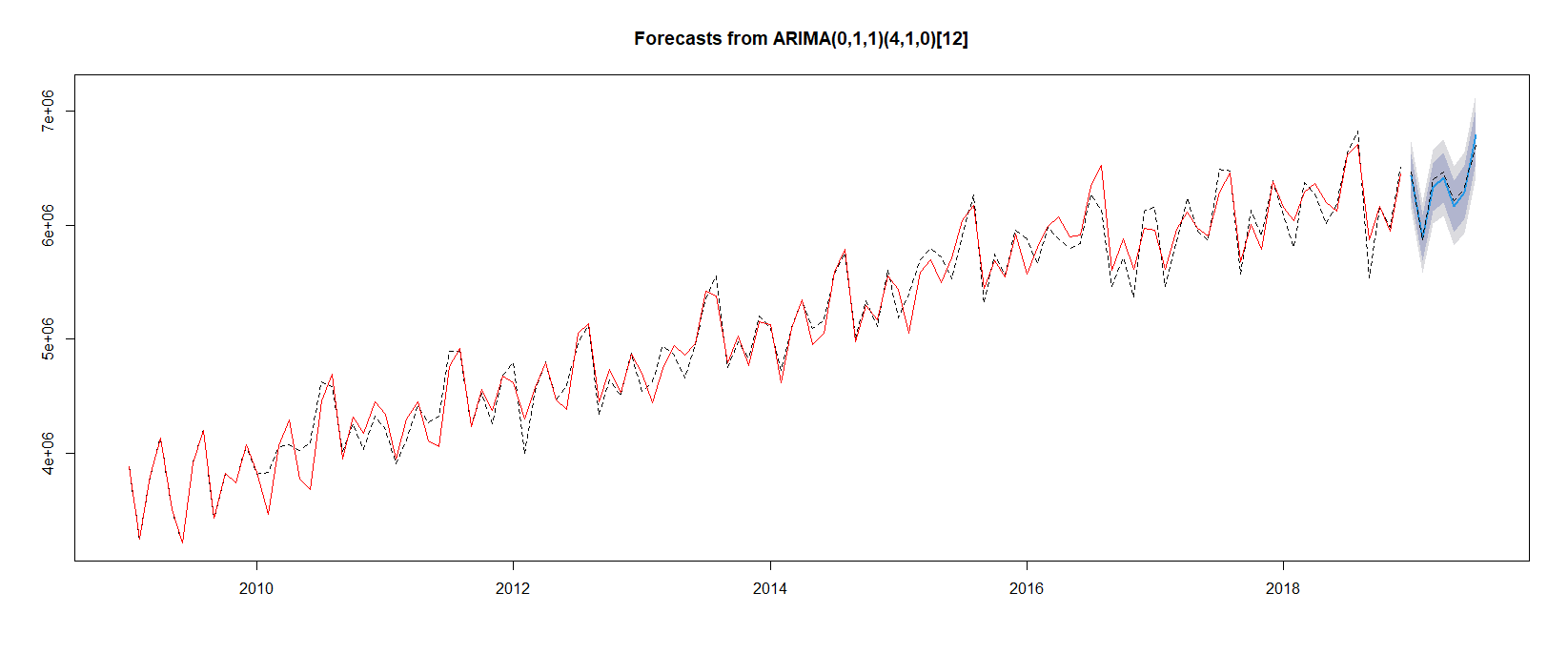}
\end{center}
   \caption{Forecast for 7 months}
\label{fig:part1_8_forecast7}
\end{figure}

\begin{table}[h]
\begin{center}
  \begin{tabular}{c|*{6}c}
  \toprule
    Month & Turth & Predicted & 95 Lower & 95 Upper & 80 Lower & 80 Upper \\
  \midrule
    Jan & 6460193 & 6431575 & 6238776 & 6624374 & 6136715 & 6726436 \\
    Feb & 5866706 & 5884746 & 5683235 & 6086257 & 5576561 & 6192930 \\
    Mar & 6396906 & 6331132 & 6121270 & 6540993 & 6010176 & 6652088 \\
    Apr & 6464336 & 6413993 & 6196101 & 6631886 & 6080755 & 6747232 \\
    May & 6209935 & 6164378 & 5938740 & 6390016 & 5819295 & 6509462 \\
    Jun & 6319690 & 6285538 & 6052412 & 6518664 & 5929003 & 6642074 \\
    Jul & 6702076 & 6787117 & 6546736 & 7027498 & 6419486 & 7154748 \\
  \bottomrule
  \end{tabular}
\end{center}
\caption{Ground truth, predicted value, and confidence intervals}
\label{table:forecast7}
\end{table}

\subsubsection{Forecast Precision Comparison}
I tried to calculate the Mean Squared Error of Model0, Model1, Model2, Model3, OverSAR, OverSMA, Model4, and Model5 mentioned before. Actually, our final choice——Model3 outperforms other models on this metric also.

This also testify that the previous decision of choosing Model3 rather than Model5, regarding Model5 is less adequate, is correct.

\begin{table}[h]
\begin{center}
  \begin{tabular}{*{8}c}
  \toprule
    Model0 & Model1 & Model2 & \textbf{Model3} & OverSAR & OverSMA & Model4 & Model5 \\
   \midrule
64532.64 & 52624.96 & 52483.63 & \textbf{51379.15} & 53735.19 & 57491.59 & 60974.08 & 61514.54 \\
    \bottomrule
  \end{tabular}
\end{center}
\caption{MSE comparison between all models fitted before}
\label{table:msecomparison}
\end{table}

\subsubsection{What will happen if we forecast 12 periods}

\begin{figure}[h]
\begin{center}
   \includegraphics[scale=0.38]{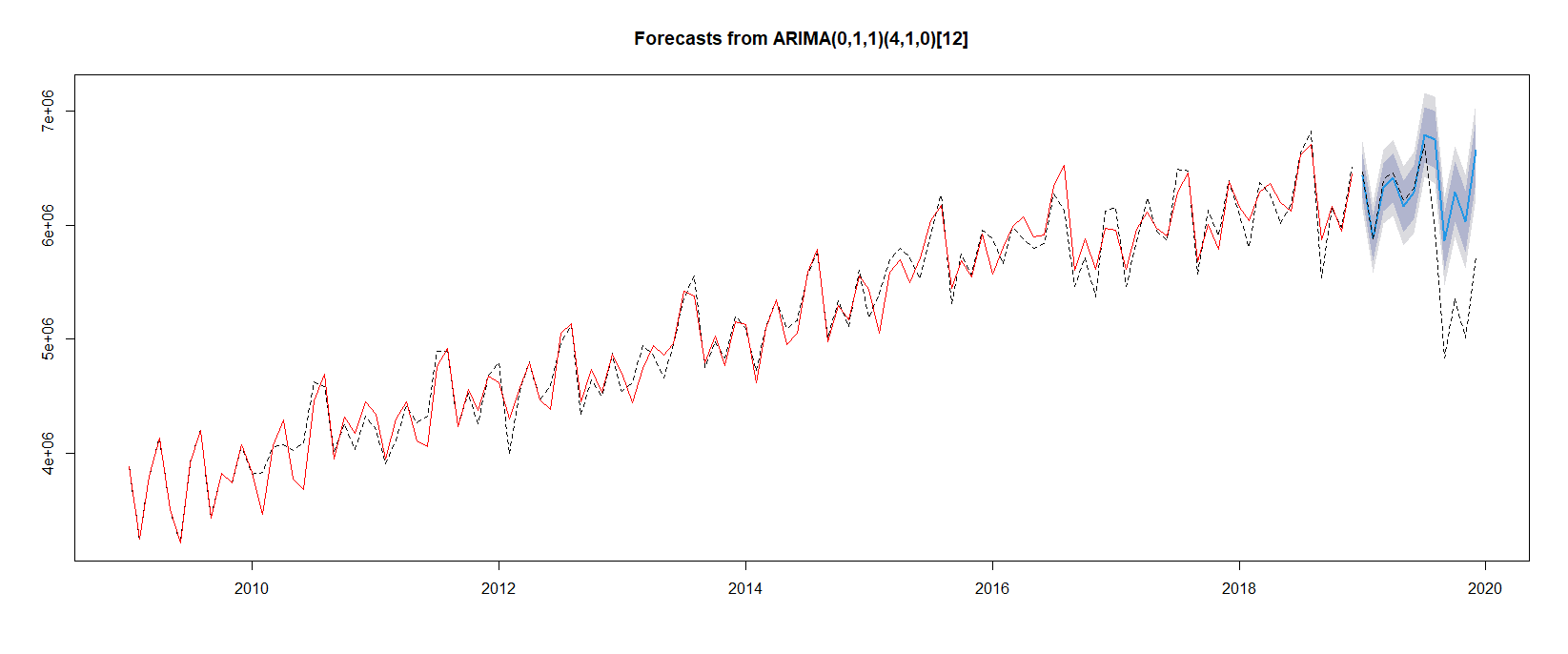}
\end{center}
   \caption{Forecast for 12 months}
\label{fig:part1_9_forecast12}
\end{figure}

In the previous forecasting, the social environment is mainly stable. From June 2020, political storm is coming, the arrival and departure air passengers are both affected greatly by the unstable social situation. If we use the selected sparse seasonal ARIMA model to predict the air traffic from August 2020 to December 2020, as shown in Figure \ref{fig:part1_9_forecast12}, the predicted value seem to be parallel with the true value but over estimated by a large margin.

Obviously, the time-series model cannot adjust itself when a major social event happened. The model will predict as if the society is developing as usual. This bring us an insight that we can use the time-series model fitted before some event happened to estimate the margin or effect caused by that event. 

\section{Quantify the Effect of Pandemic} \label{section:3}
In this section, we fit the Model3 with data from (2009,1) to (2021,1), kicking out the effect of political storm. Then, we'll use it to predict the air traffic value of the remaining eleven months in 2020.

\subsubsection{Refitting the Final Model}

The MLE of our model on new data is summarized in table \ref{table:covid_model}.

\begin{table}[h]
	\caption{Fitted Model2 and Model3}
	\centering
	\begin{tabular}{*{6}{c}}
		\toprule
		& ma1 & sar1 & sar2 & sar3 & sar4\\
		\midrule
        coefficient & -0.3925 & -0.6385 & -0.4025 & 0 & -0.2848\\
        standard error & 0.0924 &  0.0975 &  0.0941 & 0 & 0.0895\\
        \midrule
        & AIC & 3257.33 & BIC & 3271.27 & \\
		\bottomrule
	\end{tabular}
	\label{table:covid_model}
\end{table}

\subsubsection{Forecast and Quantification of Loss}

\begin{figure}[h]
\begin{center}
   \includegraphics[scale=0.38]{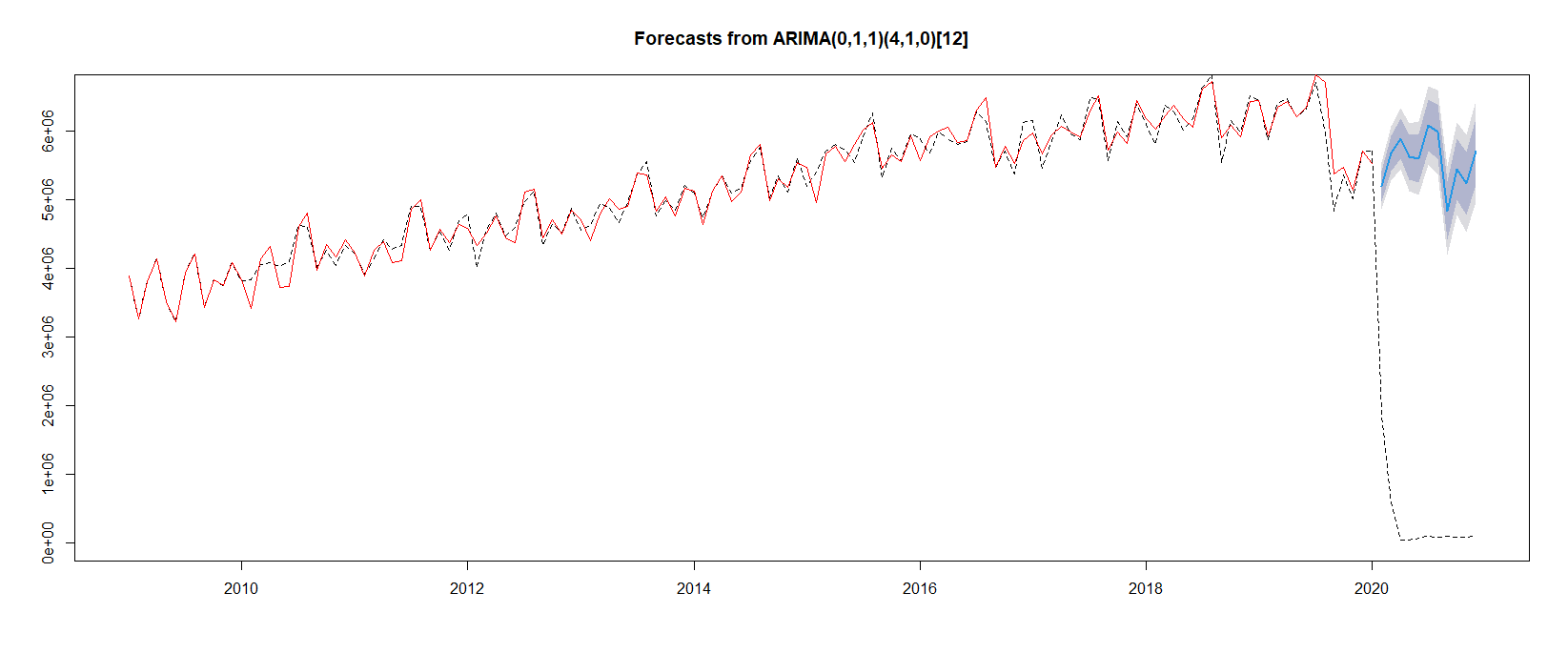}
\end{center}
   \caption{Application: Forecast for estimation}
\label{fig:part2_1_covidforecast}
\end{figure}

I take the difference between predicted value and true value as predicted loss. The rate shows that COVID-19 made loss of 95$\%$ of air passengers in the late eleven months of 2020. And the most severe month was April, with a lost of $99.95\%$ air traffic as it should have without COVID-19.

\begin{table}[h]
\begin{center}
  \begin{tabular}{c|*{6}c}
  \toprule
    Month & Turth & Predicted & Predicted Loss & 1-Loss rate \\
  \midrule
    Feb & 1877718 & 5191726 & 5683235 & 0.361675 \\
    Mar & 575825 & 5673103 & 6121270 & 0.10150 \\
    Apr & 31739 & 5883236 & 6196101 & 0.00539 \\
    May & 37423 & 5616262 & 5938740 & 0.00666 \\
    Jun & 59199 & 5597481 & 6052412 & 0.01058 \\
    Jul & 96028 & 6077110 & 6546736 & 0.01580 \\
    Aug & 83807 & 5986478 & 6238776 & 0.01400 \\
    Sep & 99805 & 4833198 & 5938740 & 0.02065 \\
    Oct & 79360 & 5442955 & 6052412 & 0.01458 \\
    Nov & 81001 & 5234962 & 6546736 & 0.01547 \\
    Dec & 90531 & 5701242 & 6238776 & 0.01588 \\    
    Overall & 3112436 & 61237753 & 58125317 & 5.08$\%$ \\
  \bottomrule
  \end{tabular}
\end{center}
\caption{Estimation of the loss of air traffic due to COVID-19}
\label{table:forecast7}
\end{table}

\section{Decomposition for Explanation} \label{section:4}

\begin{figure}[h]
\begin{center}
   \includegraphics[scale=0.38]{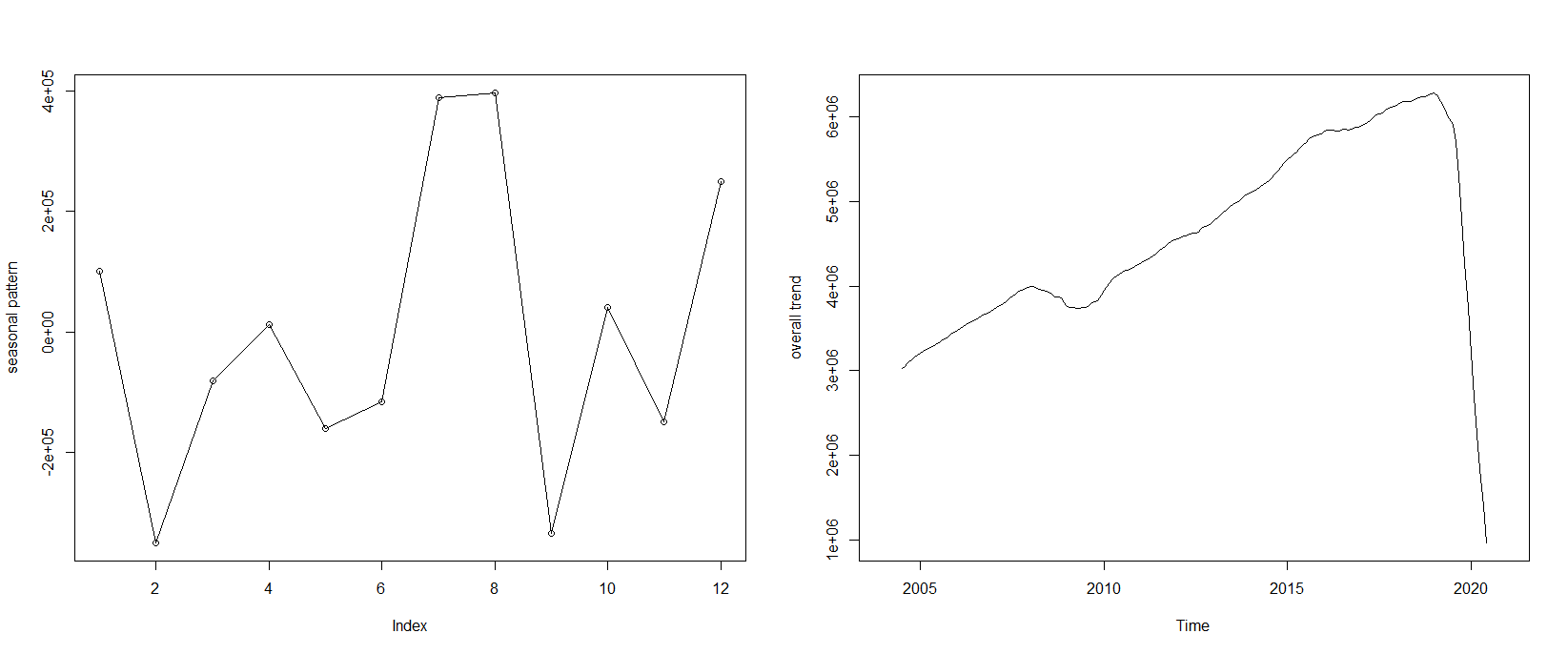}
\end{center}
   \caption{Application: Forecast for estimation}
\label{fig:part3_1}
\end{figure}

The air traffic time-series can be decomposed into three parts: trend, seasonal pattern, and random noise. Ignoring the random noise, we can plot the trend and seasonal pattern as Figure \ref{fig:part3_1}.

\subsection{Analysis of the Trend}

The level of air traffic increased as the economy developed from 2004 to 2019, except for a small drop back in 2008.

\subsection{Analysis of the Seasonal Pattern}

Generally, August and July are the peak season, following by December and January. The underlying reason might be simple: The tourism in Hong Kong is prosperous in summer, and students also flow frequently in these months. As for the winter peak, it might be partly due to the spring festival, when lots of people working or studying in Hong Kong fly back to their home city or just go for a trip.

\section{Conclusion}
Sparse Seasonal ARIMA Model3 does a great job in predicting future air traffic of Hong Kong International Airport in a stable social environment. Although it fails to react to the social change automatically, it can still be used to evaluate the effect of social change on the air traffic time-series. I believe it can be extended to other topics of economic and social statistics.

\end{document}